\documentclass[reprint,superscriptaddress,amsmath,amssymb,aps,showpacs]{revtex4-2}
\usepackage[utf8]{inputenc}
\usepackage{graphicx}
\usepackage{bm}
\usepackage{epsfig}
\usepackage{amssymb}
\usepackage{amsfonts}
\usepackage{color}
\usepackage{natbib}
\usepackage[caption=false]{subfig}

\newcommand{\la}{\langle}
\newcommand{\ra}{\rangle}

\graphicspath{{figures/}}
\usepackage{caption}

\begin{document}

\title{Quantum phase transition from a paramagnetic Anderson insulating state to a ferromagnetic many-body localized state via an intermediate ferromagnetic metallic phase}

\date{\today}

\author{Kyung-Yong Park}
\affiliation{Department of Physics, POSTECH, Pohang, Gyeongbuk 37673, Korea}

\author{Iksu Jang}
\affiliation{Department of Physics, National Tsing Hua University, Hsinchu 30013, Taiwan}

\author{Ki-Seok Kim}
\affiliation{Department of Physics, POSTECH, Pohang, Gyeongbuk 37673, Korea}
\affiliation{Asia Pacific Center for Theoretical Physics (APCTP), Pohang, Gyeongbuk 37673, Korea}

\email[Kyung-Yong Park: ]{parkky@postech.ac.kr}
\email[Ki-Seok Kim: ]{tkfkd@postech.ac.kr}

\begin{abstract}
Effects of electron correlations on Anderson insulators have been one of the central themes for recent two decades, suggesting that the Anderson insulating phase turns into a novel insulating state referred to as many body localization (MBL). However, the role of spin degrees of freedom in this dynamical phase transition still remains unclarified as a function of the interaction strength. In this study, we perform real-space spin-resolved Hartree-Fock-Anderson simulations to investigate metal-insulator transitions above a critical disorder strength in three spatial dimensions, where all single-particle states are Anderson-localized without interactions. Here, relatively weak correlations below the Mott regime are taken into account in the mean-field fashion but disorder effects are introduced essentially exactly. We find two types of single-particle mobility edges, where the multifractal spectrum of the interaction-driven low-energy mobility edge deviates from that of the high-energy one smoothly connected with the multifractal spectrum of the metal-insulator transition without interactions. We show that the weakly interacting insulating phase remains to be a paramagnetic Anderson insulating state up to the temperature of the order of the band width. On the other hand, we uncover that the relatively strongly interacting insulating phase still below the Mott regime is ferromagnetic, which turns into a ferromagnetic metallic state at a critical temperature much lower than the order of the bandwidth. Based on all these results, we propose a quantum phase transition from a paramagnetic Anderson insulating state to a ferromagnetic MBL insulating phase via an intermediate ferromagnetic metallic state, which intervenes between these two insulators at the Fermi energy.
\end{abstract}

\maketitle

\section{Introduction}

Since the discovery of Anderson localization referred to as ``Absence of Diffusion in Certain Random Lattices" \cite{Anderson_AL_Original}, it has been investigated the role of electron correlations in the Anderson insulating phase \cite{AL_Disorder_Interaction}. One of the fundamental questions is whether electron correlations can thermalize electron themselves to turn the Anderson insulating phase into a diffusive metallic state, increasing temperatures only without any other excitations such as phonons \cite{AL_Thermalization_I,AL_Thermalization_II,AL_Thermalization_III,AL_Thermalization_IV}. It turns out that the resulting disordered correlated system fails to self-thermalize at infinitesimal temperatures at least, referred to as many body localization (MBL), where the single-particle localized wave-function evolves into a many-particle localized state of the Fock space \cite{MBL_Confirmed_I,MBL_Confirmed_II}. Here, the so called many body mobility edge instead of the single-particle one was proposed to characterize the phase boundary between the diffusive metallic state and the MBL insulating phase in the energy-disorder phase diagram \cite{MBL_Review_I,MBL_Review_II,MBL_Review_III}. Later, the stability of MBL has been attributed to the emergence of quasi-local integrals of motion, rigorously verified at least in one-dimensional systems \cite{MBL_Proof_1D}. In particular, this perspective asked various questions such as the existence of MBL above one spatial dimension, the stability of the many body mobility edge, and etc. \cite{Altaman_Review,MBL_Selected_Topics} Recently, quantum phase transitions between the MBL insulating phase and metallic states have been examined in a phenomenological renormalization group approach as well as numerical simulations, suggesting anomalous quantum critical scaling behaviors \cite{MBL_QPT_RG_I,MBL_QPT_RG_II,MBL_QPT_RG_III,MBL_QPT_RG_IV,MBL_QPT_RG_V}.

Although not only the fundamental concept of MBL was established but also dynamical quantum phase transitions to nearby quantum phases have been investigated extensively, the role of spin degrees of freedom in MBL of electrons has been rarely addressed explicitly and clearly. As an example for importance of spin degrees of freedom, ferromagnetic fluctuations are well known to play a central role in two-dimensional metal-insulator transitions \cite{Finkelstein_RG}. In this study, we investigate the role of spin degrees of freedom in how the Anderson insulating phase evolves into the MBL insulating state, increasing effects of electron correlations. Here, we perform real-space spin-resolved Hartree-Fock-Anderson simulations to investigate metal-insulator transitions \cite{Two_Mobility_Edges,Half_metals} above a critical disorder strength in three spatial dimensions ($W > W_c$), where all single-particle states are Anderson-localized without interactions. To determine single-particle mobility edges, we examine a scaling behavior of the inverse participation ratio (IPR) \cite{IPR_Review} as a function of the interaction strength ($U$). We find that a paramagnetic Anderson insulating state turns into a ferromagnetic metallic phase at a critical strength of interaction $(U_{c1})$. Emergence of a metallic state by electron correlations is not inconsistent with a previous self-consistent mean-field theory study \cite{CPA_plus_MFT_AL}, which focuses on the absence of spin polarization mostly. Further increase in interactions makes the ferromagnetic metallic state develop a pseudo-gap around the Fermi energy. As a result, this metallic phase evolves into another ferromagnetic insulating state, where a low-energy mobility edge appears close to the Fermi level. These two critical points are analyzed by the multifractal spectrum, which shows that the latter transition $(U_{c2})$ contains weak eigenfunction multifractal nature associated with electron interactions. In particular, the multifractal spectrum of the interaction-driven low-energy mobility edge deviates from that of the high-energy one smoothly connected with the multifractal spectrum of the metal-insulator transition without interactions.

To understand the difference between these two insulators further, we investigate the evolution of single-particle mobility edges as a function of temperatures. We show that the second ferromagnetic insulator in $U_{c2} < U \ll U_{\rm{Mott}}$ suffers an insulator-metal transition at a critical temperature $T_c$ while the first paramagnetic insulator in $0 < U < U_{c1}$ remains to be an insulator up to the order of the bandwidth. The low-energy mobility edge disappears above the critical temperature $T_c$ and the diffusive nature dominates. These two insulators can be distinguished further by the value of entanglement entropy at zero temperature \cite{MBL_Review_I,MBL_Review_II,MBL_Review_III,MBL_Proof_1D,Altaman_Review,MBL_Selected_Topics}. Although both insulators follow the area law of the entanglement entropy, the second insulator turns out to be more entangled. Therefore, we propose a quantum phase transition from a paramagnetic Anderson insulating state to a ferromagnetic MBL insulating phase via an intermediate ferromagnetic metallic state, which intervenes between these two insulators at the Fermi energy.

Before going further, we would like to give several remarks. First, the Hartree-Fock-Anderson simulation method cannot be applied to the Mott regime. It turns out that it is certainly below the Mott regime the quantum phase transition
%
%
to occur as a function of the interaction parameter $U$. We refer this discussion to our recent study \cite{Half_metals}, which justifies the Hartree-Fock-Anderson simulation method. Second, the Hartree-Fock-Anderson simulation method takes into account electron correlations in the one-loop level (mean-field) but disorder effects essentially exactly. We recall the Finkelstein's nonlinear $\sigma-$model approach \cite{Finkelstein_RG}, where both electron correlations and disorder effects are introduced up to the two-loop order. We point out that the Finkelstein's nonlinear $\sigma-$model approach has difficulties in application to an insulating phase while the Hartree-Fock-Anderson simulation method works well in the insulating state. Third, the Hartree-Fock-Anderson simulation method considers only the single-particle dynamics, which cannot determine the many body mobility edge \cite{MBL_Review_I,MBL_Review_II,MBL_Review_III,MBL_Proof_1D,Altaman_Review,MBL_Selected_Topics}. However, we propose that the eigenfunction multifractal spectrum distinguishes the MBL insulating phase from the Anderson insulating state. The logarithmic increase of the entanglement entropy with respect to time has been suggested to be a fingerprint of the MBL insulating state \cite{MBL_Review_I,MBL_Review_II,MBL_Review_III,MBL_Proof_1D,Altaman_Review,MBL_Selected_Topics}. On the other hand, the energy-level statistics \cite{Level_Statistics_Review} cannot differentiate the MBL insulating phase from the Anderson insulating state. Here, we suggest the eigenfunction multifractal spectrum as a novel measure towards the MBL insulating phase. Fourth, the quantum phase transition from the paramagnetic Anderson insulating state to the ferromagnetic metallic phase results from enhancement of effective interactions between localized electrons, which may be regarded as the Stoner instability \cite{Stoner_Original,Stoner_Instability_Textbook}. On the other hand, the quantum phase transition from the ferromagnetic metallic phase to the ferromagnetic MBL insulating state originates from the formation of a pseudo-gap near the Fermi energy. This reentrant insulating behavior shows the role of spin degrees of freedom in these dynamical phase transitions.

\section{Model and Method}

\subsection{Model}

We consider the following effective model Hamiltonian on a cubic lattice, where the non-interaction part is
\begin{align}
    H_0 = - t \sum_{\langle i j \rangle} \sum_{\sigma} c_{i \sigma}^{\dagger} c_{j \sigma} + h. c. + \sum_{i} \sum_{\sigma} (\epsilon_i - \mu) c_{i \sigma}^{\dagger} c_{i \sigma} ,
\end{align}
and the interaction part is
\begin{align}
    H_I = \frac{1}{2} \sum_{i} \sum_{j} \sum_{\sigma \sigma'} c_{i \sigma}^{\dagger}c_{i \sigma} U_{ij}^{\sigma\sigma'} c_{j \sigma'}^{\dagger}c_{j \sigma'}.
\end{align}
Here, $t$ is the hopping integral between nearest neighbor sites $\langle i j \rangle$, and $\mu$ is the chemical potential. In this study we set $t = 1$ as the unit of energy and focus on the case of half filling. $\epsilon_i$ is a random potential, uniformly distributed in $[- W, W]$. It is well known that the Anderson transition occurs at $W_{c} = 8.25$ without electron correlations \cite{AMIT_Critical_Disorder_Strength}. Here, we set $W = 10 > W_c$. The interaction coefficient is given by  $U_{ij}^{\sigma\sigma'} = U \delta_{- \sigma \sigma'} \delta_{ij}$. $\sigma$ represents $+$ ($-$) for spin $\uparrow$ ($\downarrow$) electrons.

\subsection{Method I: Eigenfunction multifractal analysis}

To analyze a scaling behavior of the IPR, we partition a system of size $L^d$ into $(L/l)^d$ boxes. Each box has a probability density, given by
\begin{align}
    \mu_k \equiv \sum_{j \in \rm{box\ k}} | \Psi_j|^2 ,
\end{align}
where $\Psi_{j}$ is an eigenfunction of the effective Hamiltonian $H=H_0+H_I$ at position $j$ inside the box number $k$. In addition, we introduce a generalized IPR by summing over the eigenfunction moment of each box as
\begin{align}
    R_q=\sum_{k} \mu_k^q .
\end{align}
To get an intuitive picture, one may regard $\mu_{k}$ as $e^{- H_{k}}$ and $q$ as $\beta$, where $H_{k}$ corresponds to an effective Hamiltonian with an index $k$ and $\beta$ is an inverse temperature. Then, $R_{q}$ is analogous to the canonical partition function, given by summation of the Boltzmann factor.

Following this analogy further, we may introduce the probability density of the $q^{th}$ moment of the box number $k$ as
\begin{align}
    x_k=[\mu_k]^q/R_q .
\end{align}
This corresponds to the density matrix, given by the Boltzmann factor divided by the partition function. Now, we introduce the multifractal exponent as \cite{Chhabra1989,Janssen1994}
\begin{align}
    \alpha_q=\frac{1}{\ln \lambda} \sum_k x_k \cdot \ln \mu_k ,
\end{align}
analogous to the averaged energy, where the scaling factor is $\lambda=l/L$. Accordingly, the ``entropy" is defined as
\begin{align}
    f(\alpha_q)=\frac{1}{\ln \lambda} \sum_k x_k \cdot \ln x_k ,
\end{align}
identified with the eigenfunction multifractal spectrum.

In this study, we perform our simulations for the three-dimensional cubic lattice, varying the system size $L^{3}$ with $L = 12$, $16$, and $20$ to identify the mobility edge. We perform the disorder average for $100 \sim 1000$ disorder realizations depending on the system size.
%
%

\subsection{Method II: Entanglement entropy}

As a probe to distinguish two insulating phases, we calculate entanglement entropy. Resorting to refs. \cite{ChungPeschel,Peschel,ChungHenley}, we obtain the entanglement entropy for the subsystem $A$ as follows
\begin{align}
S_E(A)=-\sum_{\alpha}[(1-\xi_\alpha)\ln(1-\xi_\alpha)+\xi_\alpha\ln\xi_\alpha] .
\end{align}
Here, $\xi_\alpha$ is an eigenvalue of the correlation matrix given by
\begin{align}
C_{i\sigma,j\sigma'}=\frac{tr[e^{-\beta(H_{HFA}-\mu N)}c_{i,\sigma}^\dagger c_{j,\sigma'}]}{tr[e^{-\beta(H_{HFA}-\mu N)}]} .
\end{align}
$H_{HFA}$ is an effective Hartree-Fock-Anderson Hamiltonian in our simulations. $i,j$ are position indices belonging to the subsystem $A$. $\sigma,\sigma'$ are spin indices, $\beta$ is an inverse temperature, and $\mu$ is a chemical potential.

\section{Quantum phase transition from a paramagnetic Anderson insulating state to a ferromagnetic many-body localized state via an intermediate ferromagnetic metallic phase}

\subsection{Emergence of two types of mobility edges}

\begin{figure}
\includegraphics[scale=0.25]{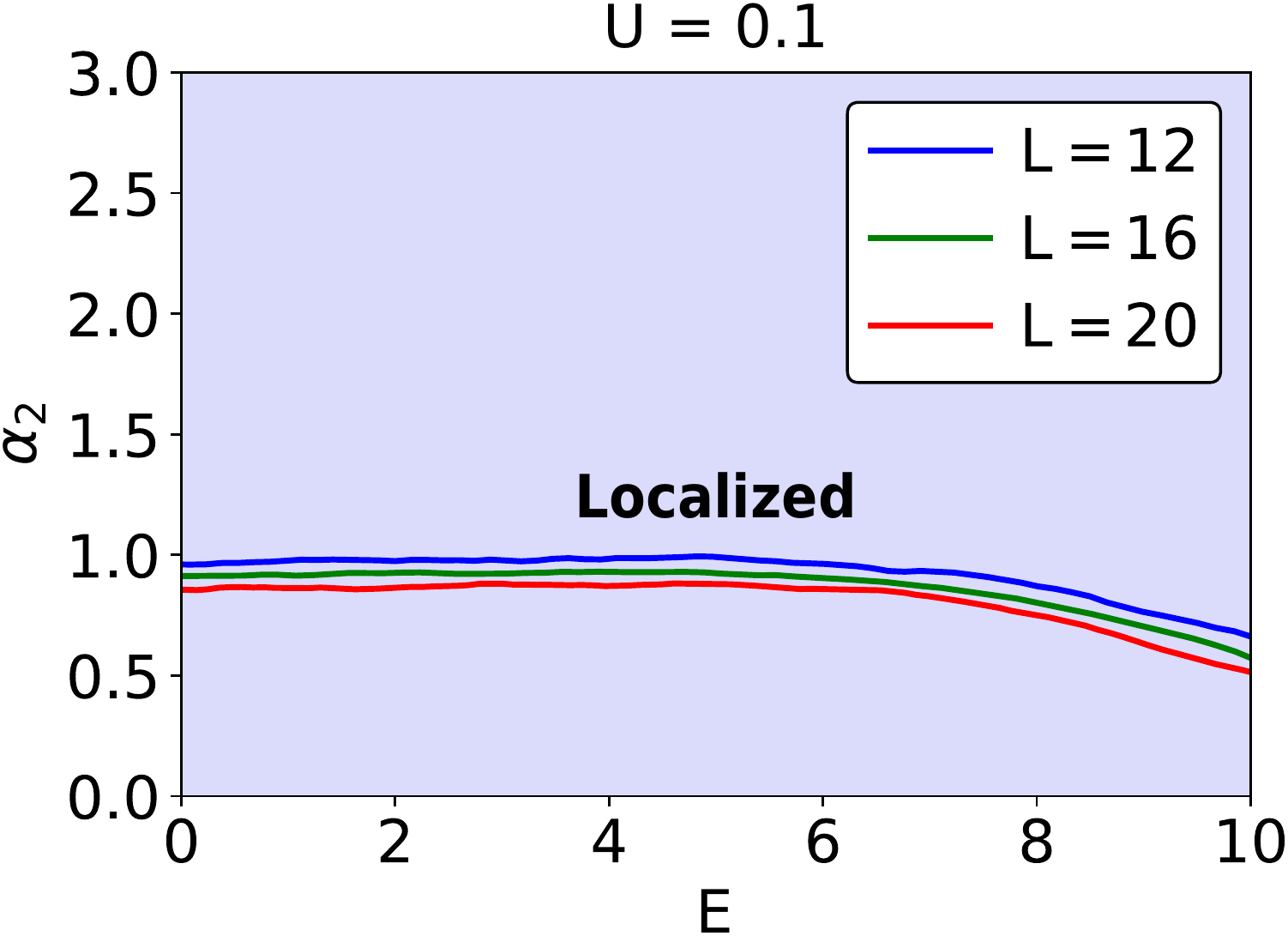}
\includegraphics[scale=0.25]{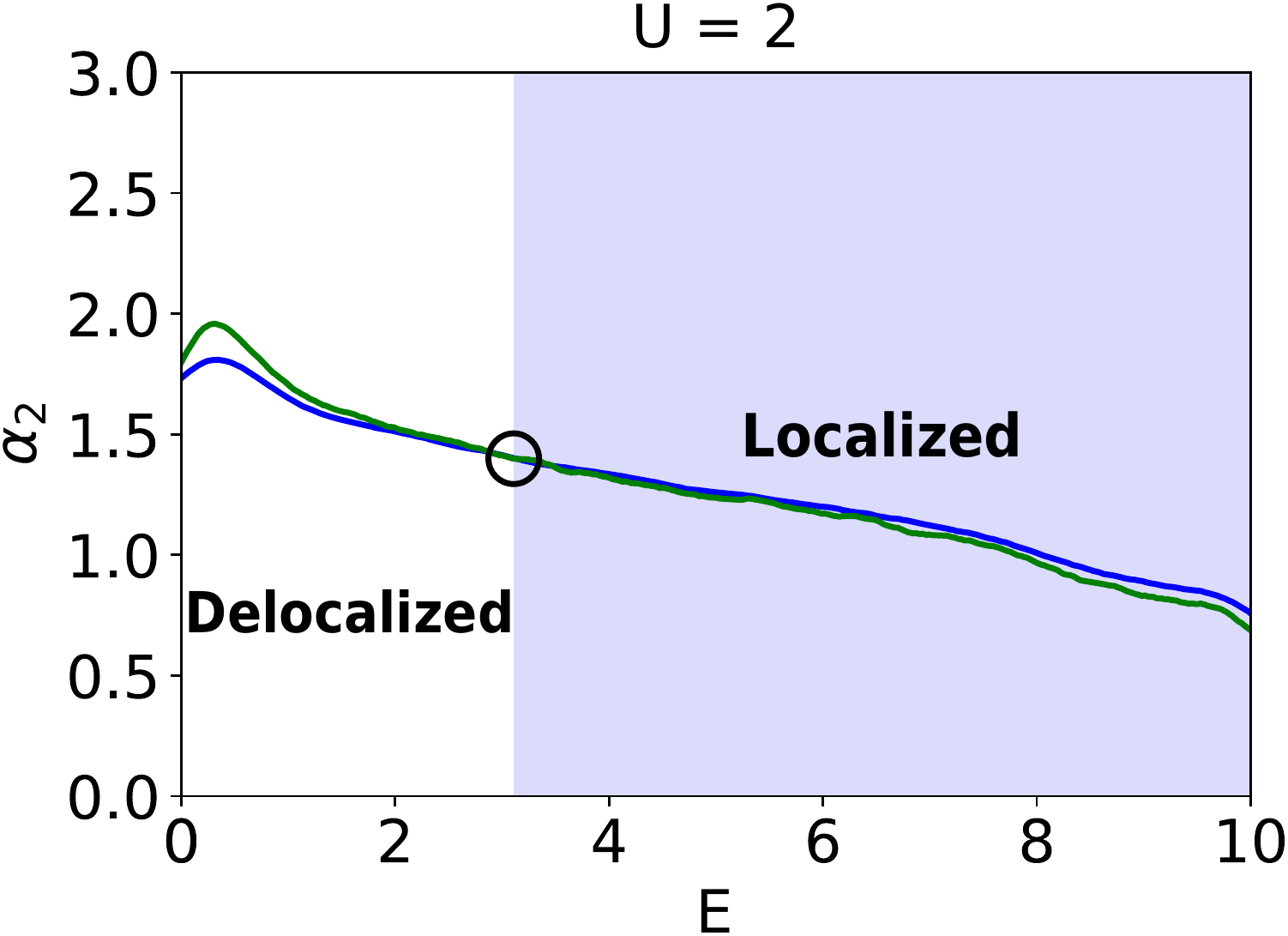}
\includegraphics[scale=0.25]{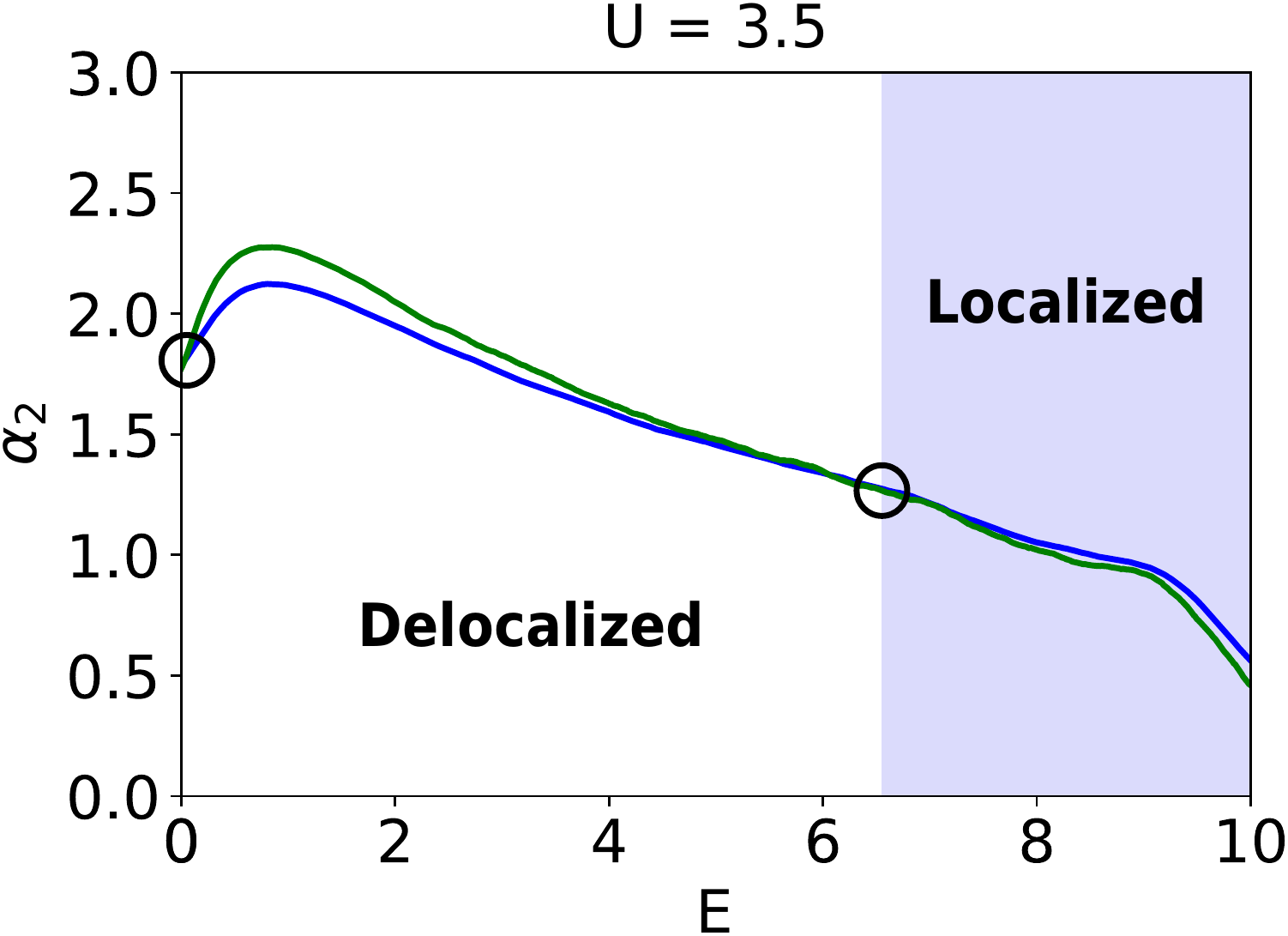}
\includegraphics[scale=0.25]{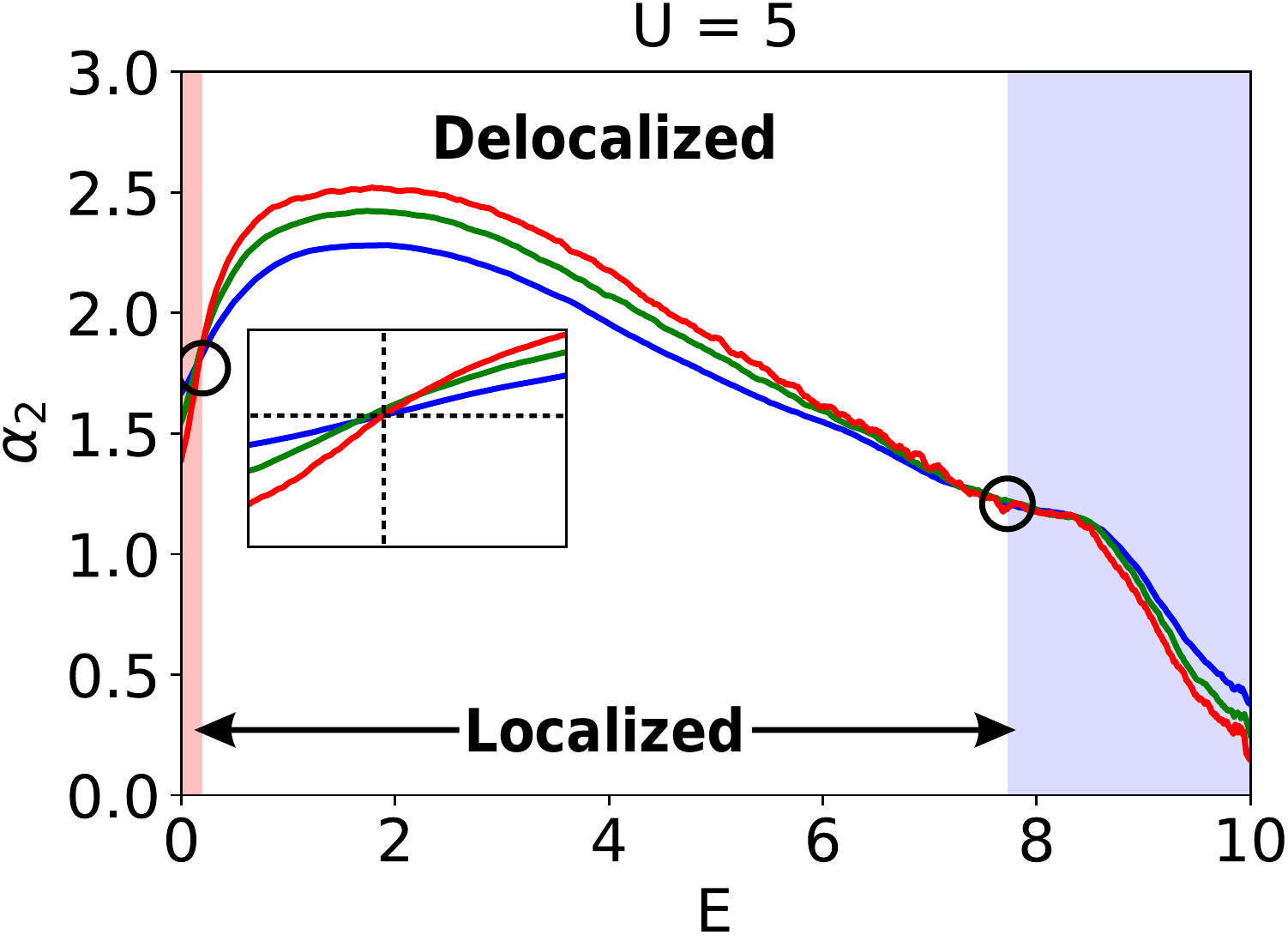}
  \caption{Scaling behavior of $\alpha_2$ for various interaction parameters of $U=0.1$, $U=2.0$, $U=3.5$, and $U=5$ at the disorder strength $W=10 > W_{c}$. Here, crossing points for three system sizes of $L = 12$, $16$, and $20$ are marked by an open circle. The inset of the last figure clarifies the crossing point near the Fermi energy. Two types of mobility edges appear to suggest insulator-metal-insulator transitions at the Fermi energy as a function of the interaction parameter.}
 \label{fig:alpha2_t0}
\end{figure}

\begin{figure}
\includegraphics[scale=0.25]{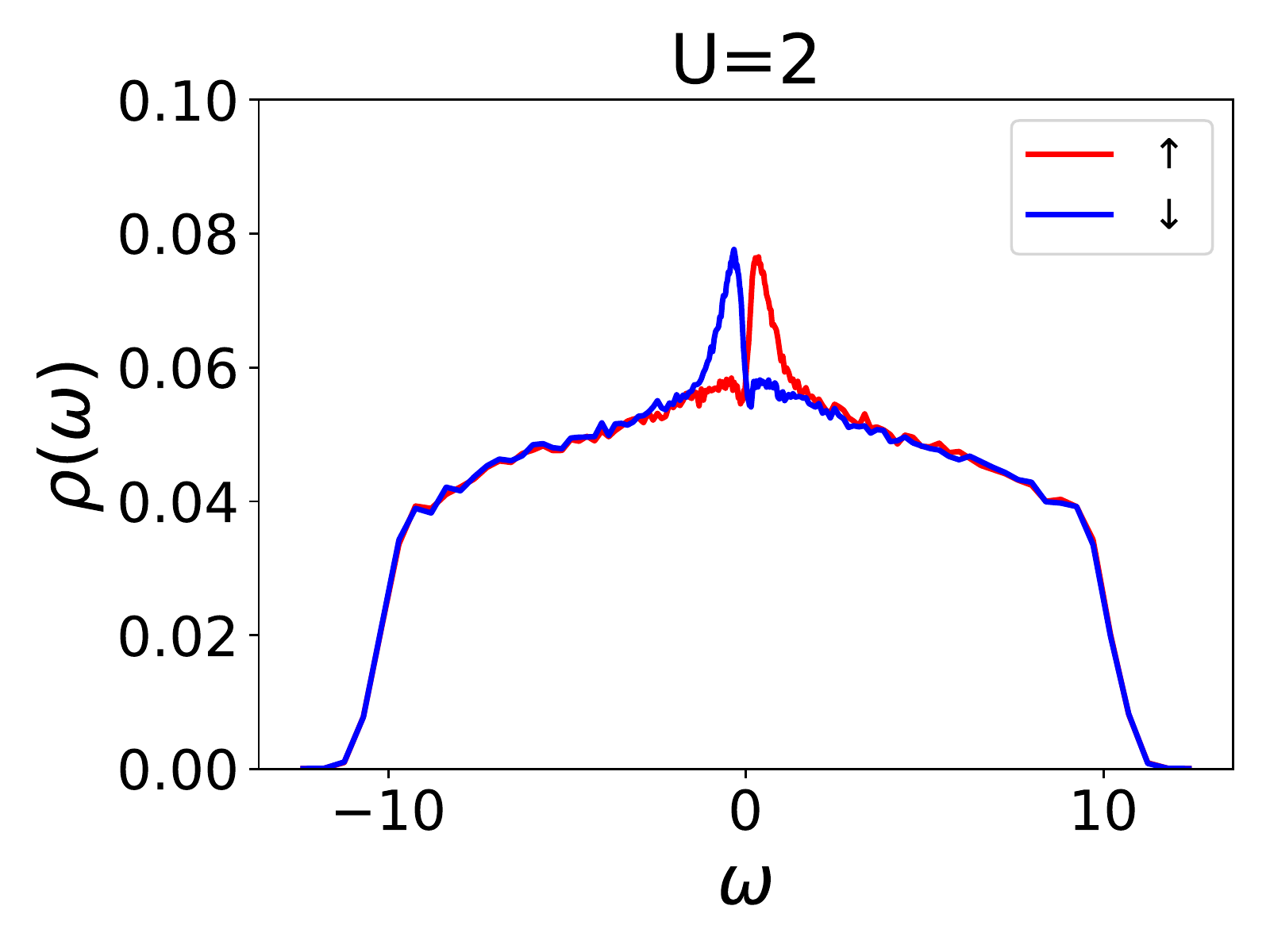}
\includegraphics[scale=0.25]{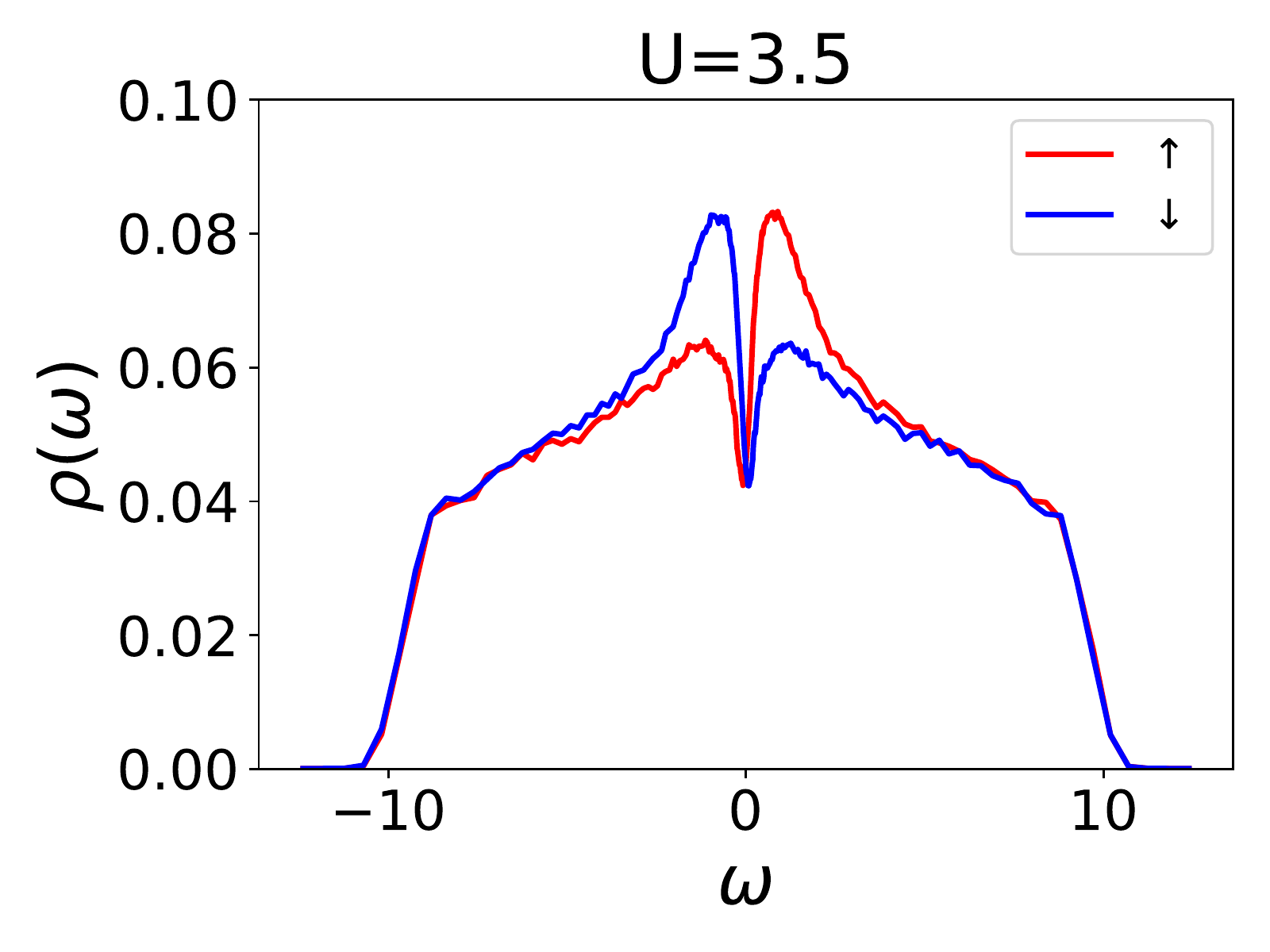}
  \caption{Density of states for interaction parameters of $U=2.0$ and $U=3.5$ at the disorder strength $W=10 > W_{c}$. The red (blue) line represents the density of states for spin $\uparrow$ ($\downarrow$) electrons. Enhancement of density of states via spin polarization is responsible for the first insulator to metal transition near the Fermi energy. Formation of a pseudogap near the Fermi energy gives rise to the second metal to insulator transition, which results from stronger spin polarization.}
 \label{fig:dos_t0}
\end{figure}

\begin{figure}
\includegraphics[scale=0.5]{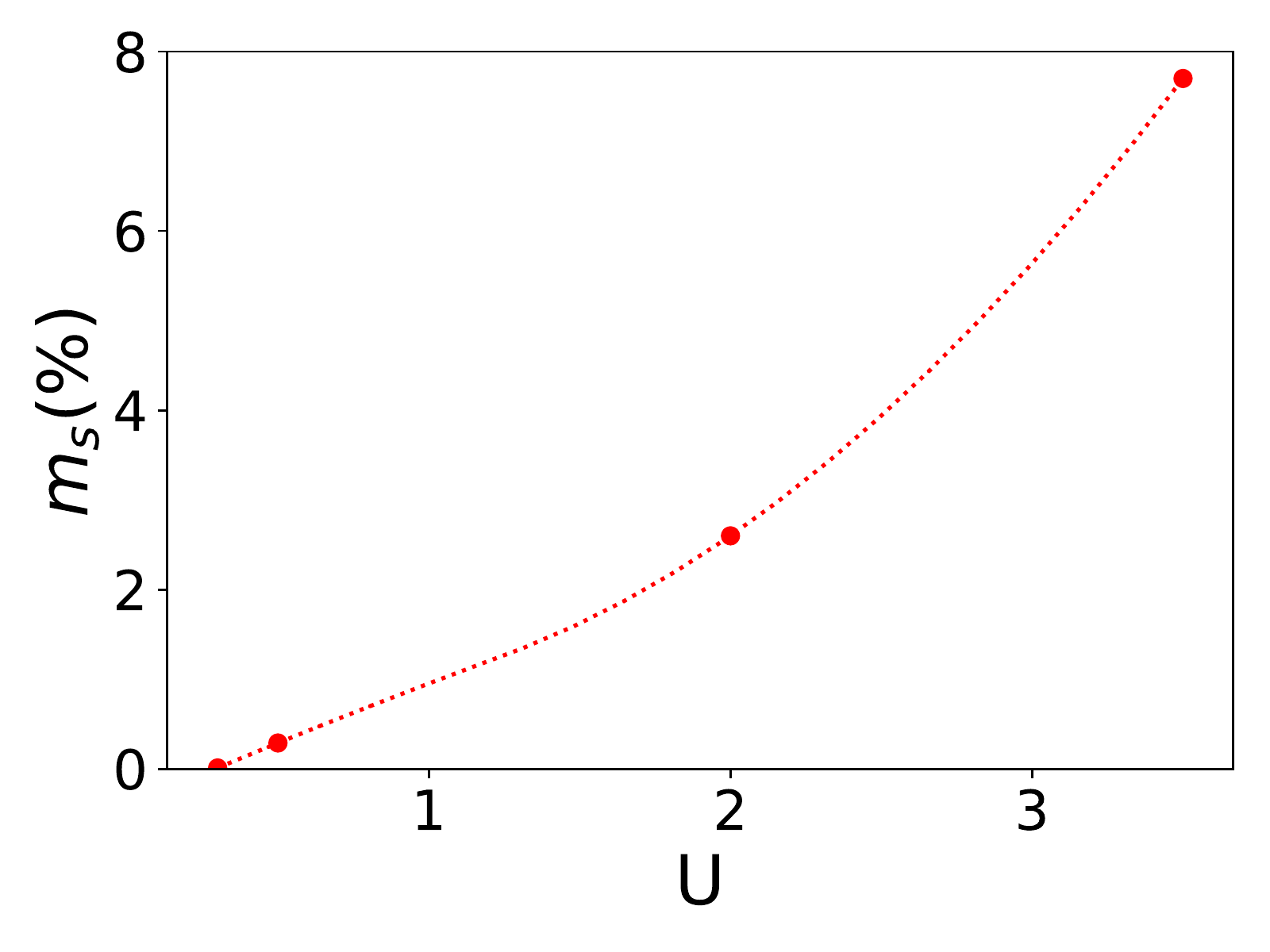}
  \caption{Magnetization $m_s = \frac{|n_{\uparrow}-n_{\downarrow}|}{n_{\uparrow}+n_{\downarrow}}$ as a function of the interaction strength $U$. Magnetization appears from $U \approx 0.3$, qualitatively similar to the critical strength that the first insulator to metal transition occurs. See the phase diagram of Fig. \ref{fig:phase1}.}
 \label{fig:mag}
\end{figure}

In the strong disorder regime of $W>W_c$, all single-particle states are Anderson-localized unless electronic correlations are taken into account. This fundamental property of the Anderson insulator still survives relatively weak interactions according to our numerical results, consistent with ref. \cite{CPA_plus_MFT_AL}. See Fig. \ref{fig:alpha2_t0} (a), which shows an insulating scaling behavior of the IPR approaching to zero as increasing the system size. All these localized states survive up to the interaction strength $U_{c1}$, from which electrons close to the Fermi energy begin to delocalize showing the opposite tendency of the scaling behavior of the IPR around the Fermi Energy. See Fig. \ref{fig:alpha2_t0} (b). This is attributed to the enhancement of density of states around the Fermi Energy, which results from spin polarization \cite{Comments_no_polarization}. See Fig. \ref{fig:dos_t0} (a), which shows appearance of a ferromagnetic metallic phase at the Fermi energy. We refer the evolution of magnetization with respect to the interaction strength to Fig. \ref{fig:mag}. Interestingly, the existence of these delocalized states is allowed until a pseudogap starts to develop at the critical point $U_{c2}$. This pseudogap behavior results from further spin polarization due to larger interactions. See Fig. \ref{fig:dos_t0} (b), where the system returns to be localized but more ferromagnetic (Fig. \ref{fig:mag}), shown in Fig. \ref{fig:alpha2_t0} (c). The low-energy mobility edge close to the Fermi energy is clearly shown when $U>U_{c2}$. See Fig. \ref{fig:alpha2_t0} (d).

\begin{figure}
\includegraphics[scale=0.5]{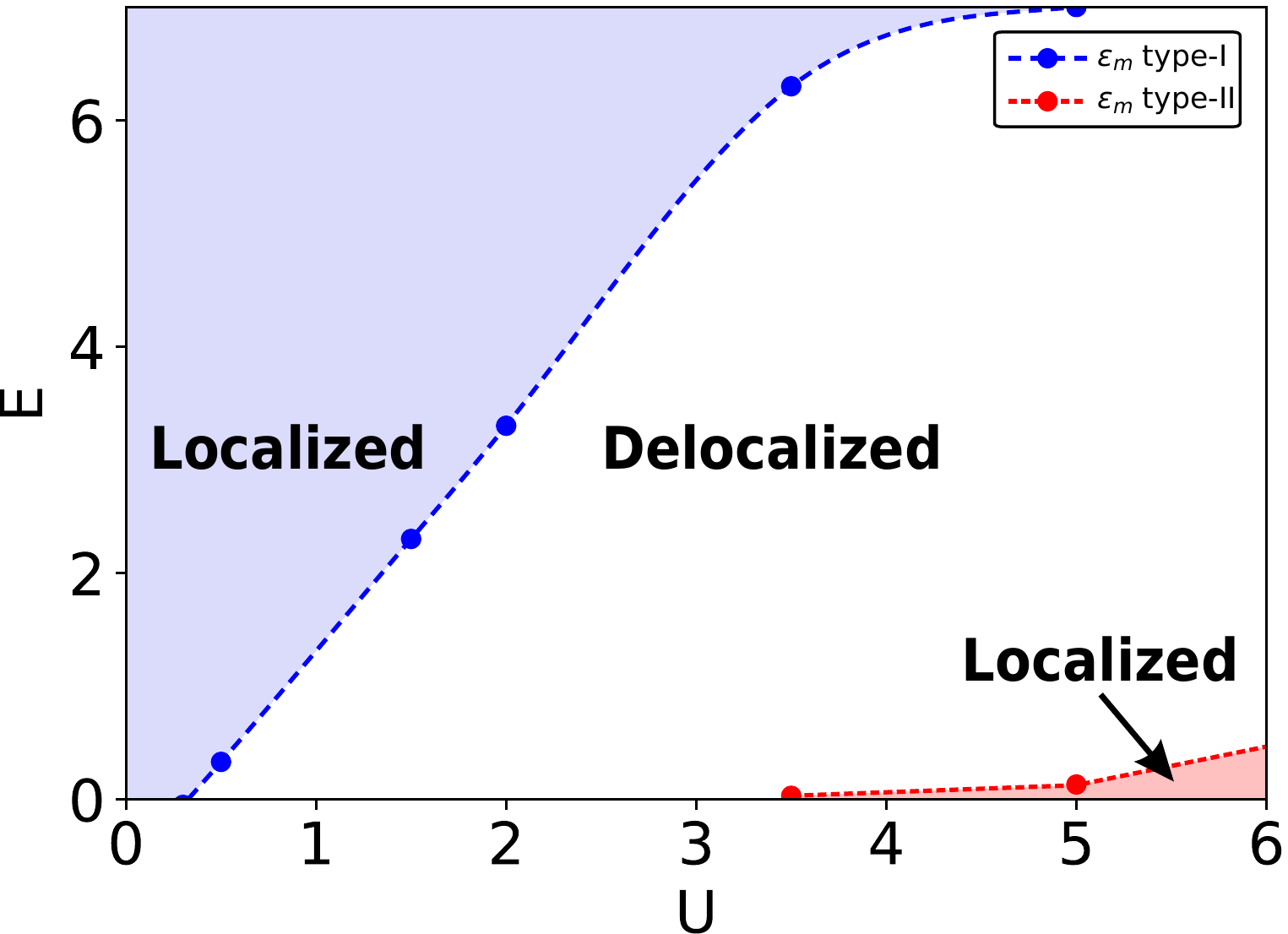}
  \caption{Phase diagram in the plane of energy and interaction strength for a strong disordered system ($W=10 > W_{c}$). Two types of insulators arise as a function of both energy and interaction, where a metallic phase intervenes between them.}
 \label{fig:phase1}
\end{figure}

To sum up, we find two types of mobility edges as a function of the interaction parameter $U$ and propose a phase diagram of Fig. \ref{fig:phase1} in the plane of energy and interaction. Here, the type I (II) mobility edge is represented by the blue (red) dot to distinguish a paramagnetic (ferromagnetic) insulating state from a ferromagnetic metallic phase, given by $U_{c1} \approx 0.3$ ($U_{c2} \approx 3.5$) at the Fermi energy. The type I mobility edge tends to shift away from the Fermi energy as increasing interactions and reaches the band edge for strong interactions ($U>U_{c2}$). If we focus on the system with $U>U_{c2}$, two types of mobility edges are shown in the single-particle spectrum at the same time. Appearance of these two types of mobility edges has been investigated in the weak localization regime ($W<W_c$) \cite{Two_Mobility_Edges,Half_metals}. This previous study told us that there exists a diffusive metallic phase between the two critical points ($U_{c1}<U<U_{c2}$), which separates such two insulators.

\begin{figure}
\includegraphics[scale=0.25]{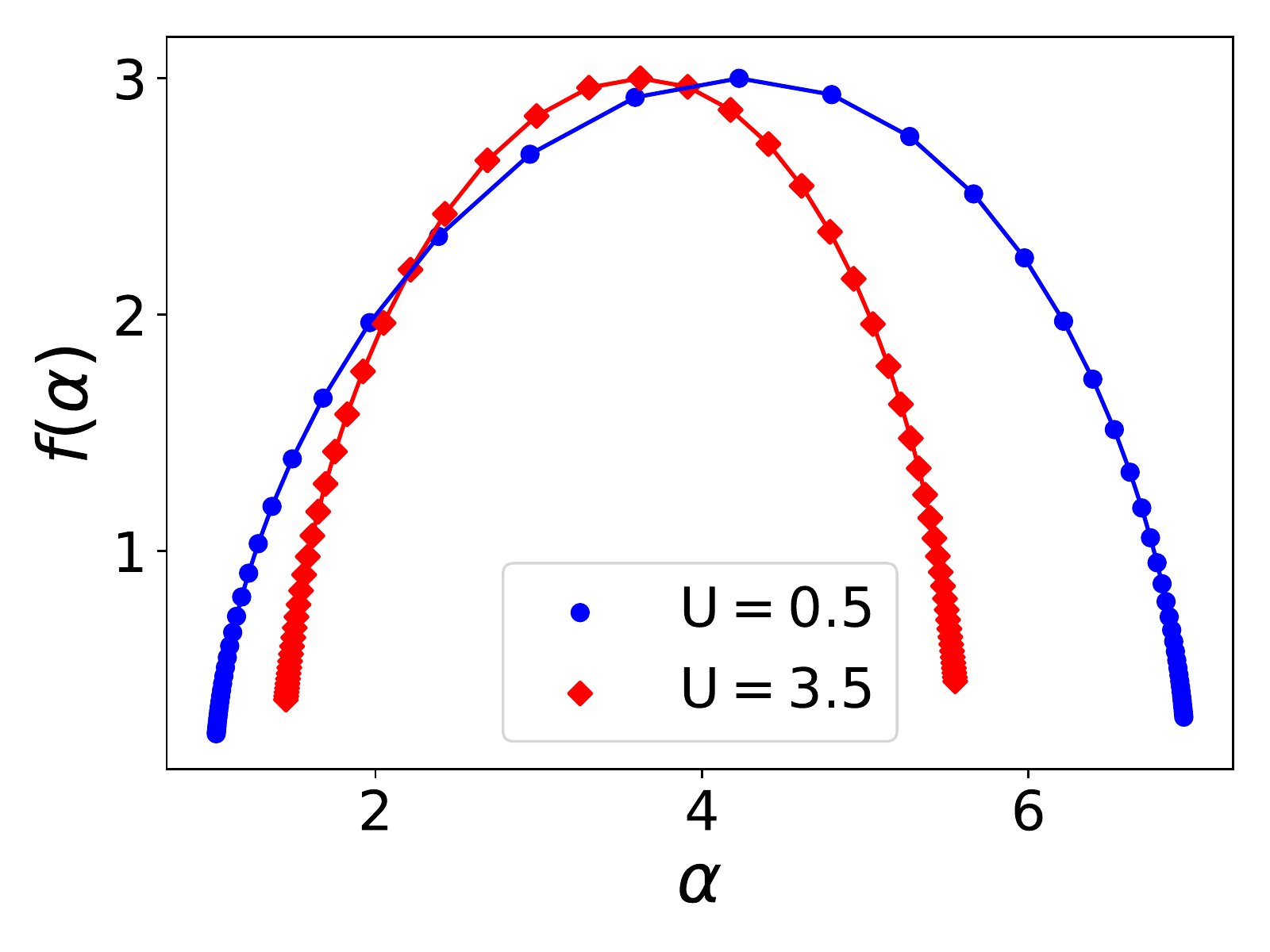}
\includegraphics[scale=0.25]{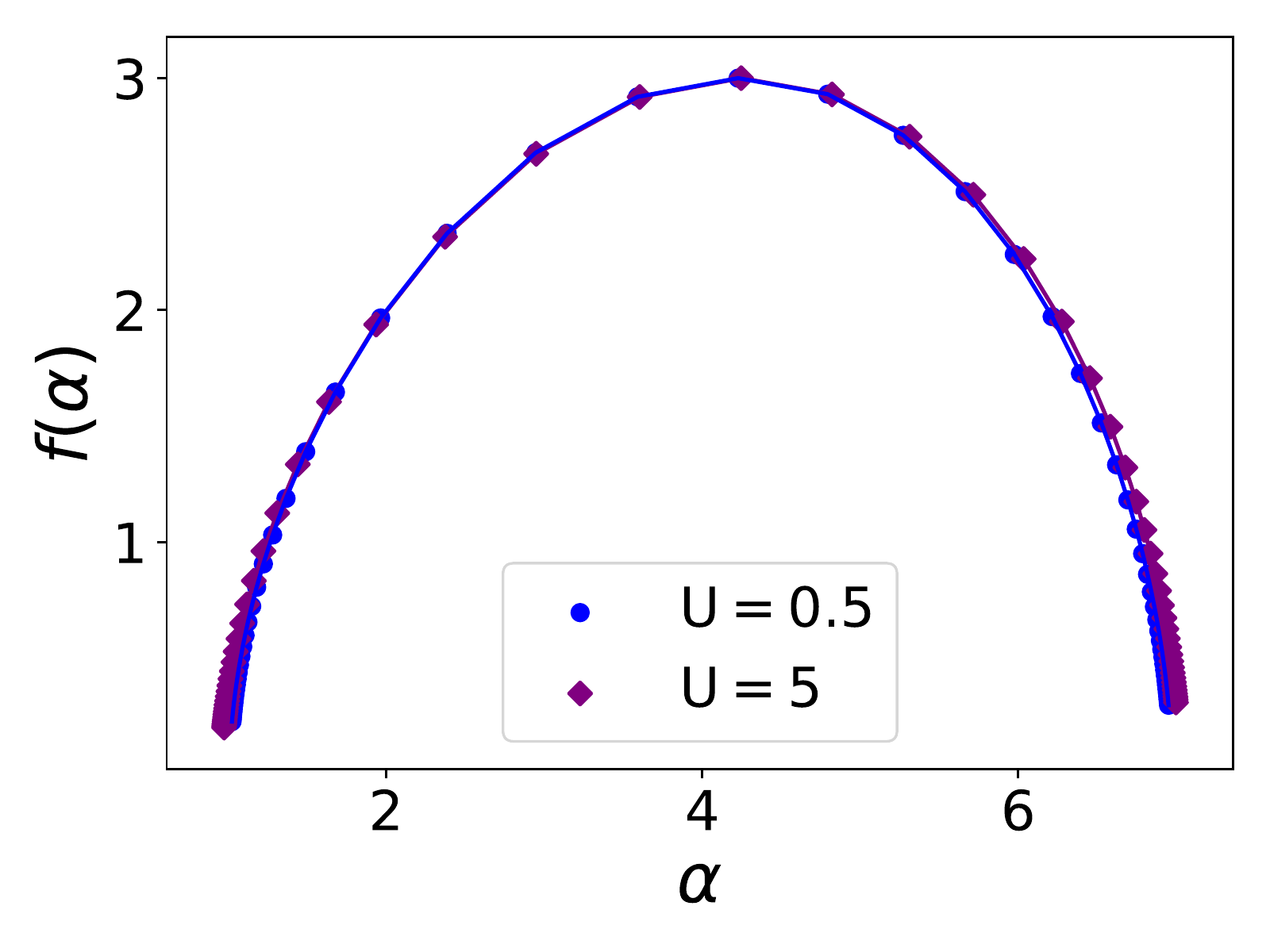}
\includegraphics[scale=0.25]{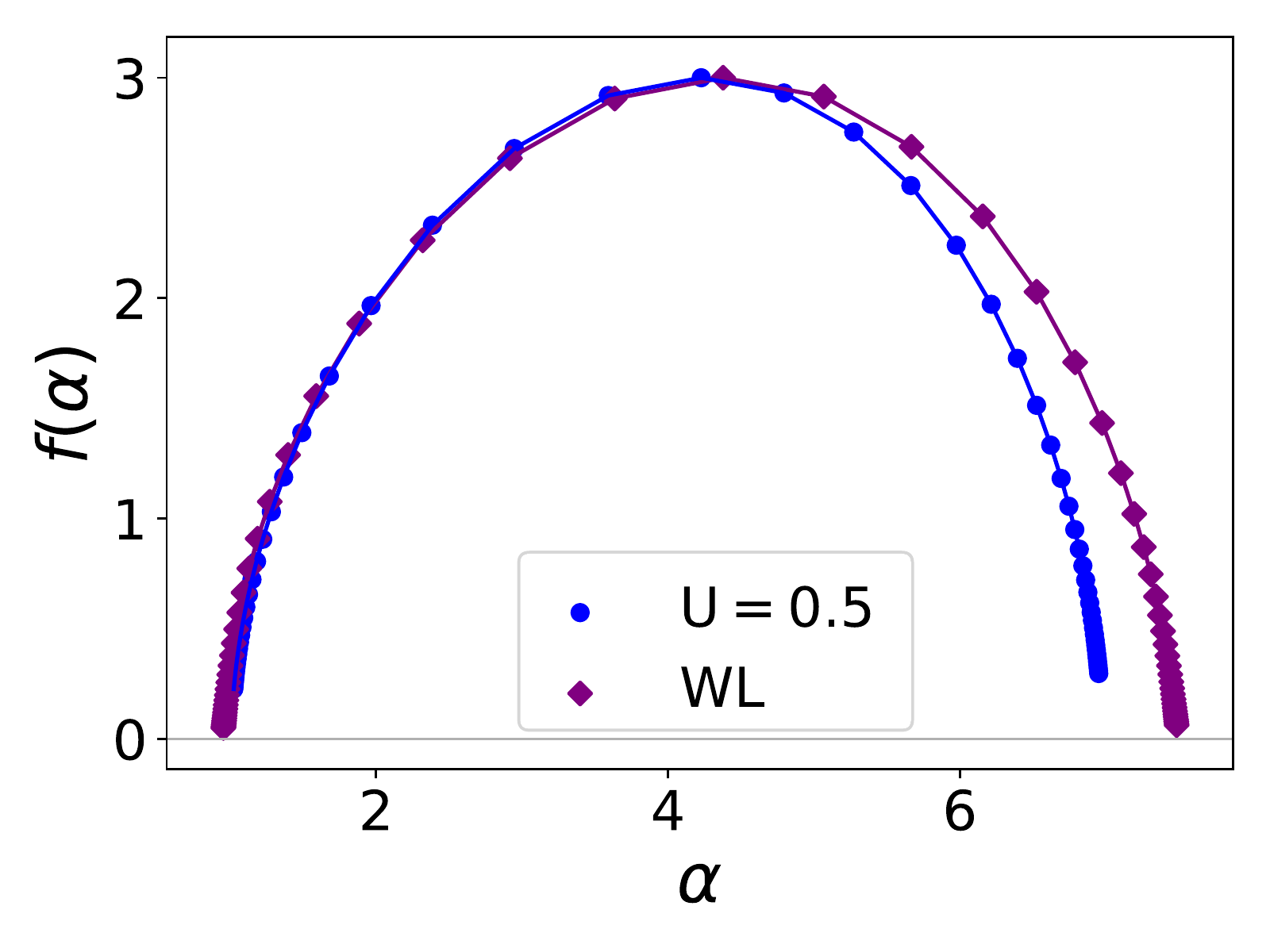}
\includegraphics[scale=0.25]{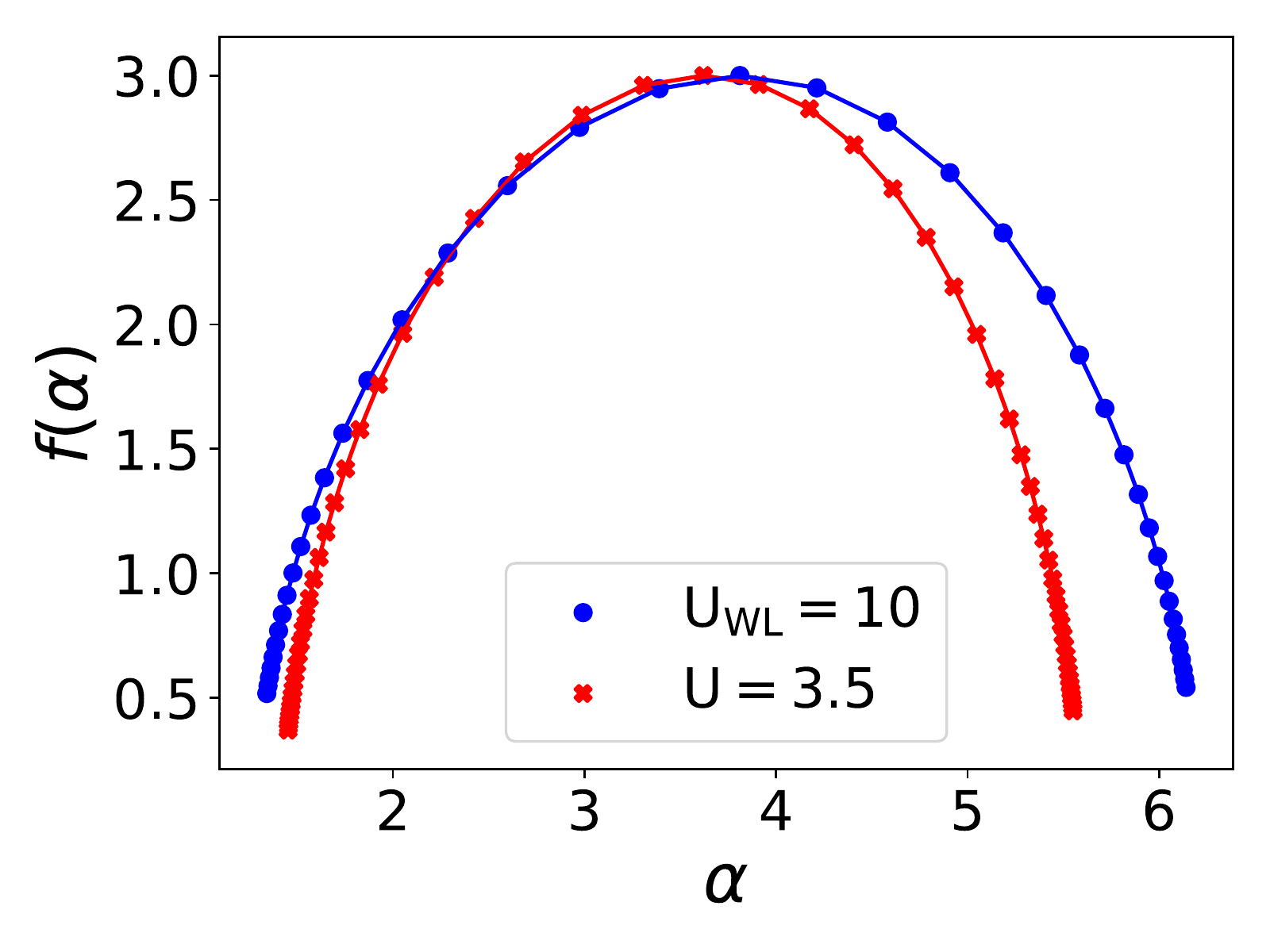}
  \caption{Multifractal spectrum to show two types of mobility edges. Here, we call them type I and type II, respectively. (a) Comparison between the multifractal spectrum of the type I mobility edge of $U=0.5$ at $E_m=0.3$ (blue dot) and that of the type II mobility edge of $U=3.5$ at $E_m=0.03$ (red diamond). Although both mobility edges lie in the low-energy region of the Fermi energy, the former deviates the latter clearly, indicating two kinds of universality classes for quantum phase transitions. (b) Comparison between the multifractal spectrum of the low-energy mobility edge of $U=0.5$ at $E_m=0.3$ (blue dot) and that of the high-energy mobility edge of $U=5$ at $E_m=7.5$ (purple diamond). This confirms that these two metal-insulator transitions belong to the same universality class, referred to as the type I mobility edge. (c) Comparison between the multifractal spectrum of the type I mobility edge of $U=0.5$ at $E_m=0.3$ (blue dot) and that of the high-energy mobility edge in the weak-disorder region ($W = 7< W_{c}$) of $U = 10$ at $E_m = 9.00$ (purple diamond).
  %
  %
  (d) Comparison between the multifractal spectrum of the type II mobility edge of $U=3.5$ at $E_m=0.03$ (red diamond) and that of the low-energy mobility edge in the weak-disorder region ($W =7 < W_{c}$) of $U = 10$ at $E_m =0.03 $ (blue dot).
  %
  %
  See the text for more details on (c) and (d).}
 \label{fig:f_a}
\end{figure}

To understand the nature of these two kinds of quantum phase transitions at $U_{c1}$ and $U_{c2}$, respectively, we obtain the multifractal spectrum for the type I mobility edge of $U=0.5$ at $E_m=0.3$ (blue dot) and compare this multifractal spectrum with that for the type II mobility edge of $U=3.5$ at $E_m=0.03$ (red diamond). See Fig. \ref{fig:f_a} (a), where the multifractal spectrum for the mobility edge close to $U_{c2}$ shows weak multifractality. Although both mobility edges lie in the low-energy region of the Fermi energy, the former deviates the latter clearly, indicating two kinds of universality classes for quantum phase transitions. Next, we compare the multifractal spectrum of the low-energy mobility edge of $U=0.5$ at $E_m=0.3$ (blue dot) with that of the high-energy mobility edge of $U=5$ at $E_m=7.5$ (purple diamond). See Fig. \ref{fig:f_a} (b). This confirms that these two metal-insulator transitions belong to the same universality class, referred to as the type I mobility edge and marked in the phase diagram of Fig. \ref{fig:phase1}. To reveal the nature of the type I mobility edge, we compare the multifractal spectrum of the type I mobility edge of $U=0.5$ at $E_m=0.3$ (blue dot) with that of the high-energy mobility edge in the weak-disorder region ($W = 7< W_{c}$) of $U = $ at $E_m = $ (purple diamond). See Fig. \ref{fig:f_a} (c). It turns out that the multifractal spectrum of the type I mobility edge belongs to the same universality class as that of the Anderson transition without electron correlations. Previously, we showed the existence of two types of mobility edges in the metallic phase near the Fermi energy in the weak-disorder region ($W = 7< W_{c}$) of $U = $ \cite{Two_Mobility_Edges,Half_metals}. The multifractal spectrum of the high-energy mobility edge in the weak-disorder region ($W = 7< W_{c}$) was shown to be in the same universality class of the Anderson transition in our previous study. Finally, to reveal the nature of the type II mobility edge, we further compare the multifractal spectrum of the type II mobility edge of $U=3.5$ at $E_m=0.03$ (red diamond) with that of the low-energy mobility edge in the weak-disorder region ($W = 7< W_{c}$) of $U = 10$ at $E_m = $ (blue dot). See Fig. \ref{fig:f_a} (d). It is not completely confirmed that the multifractal spectrum of the type II mobility edge is smoothly connected to that of the low-energy mobility edge in the weak-disorder region.

\subsection{Existence of an insulator to metal transition at a critical temperature in the type II insulator}

\begin{figure}
\includegraphics[scale=0.25]{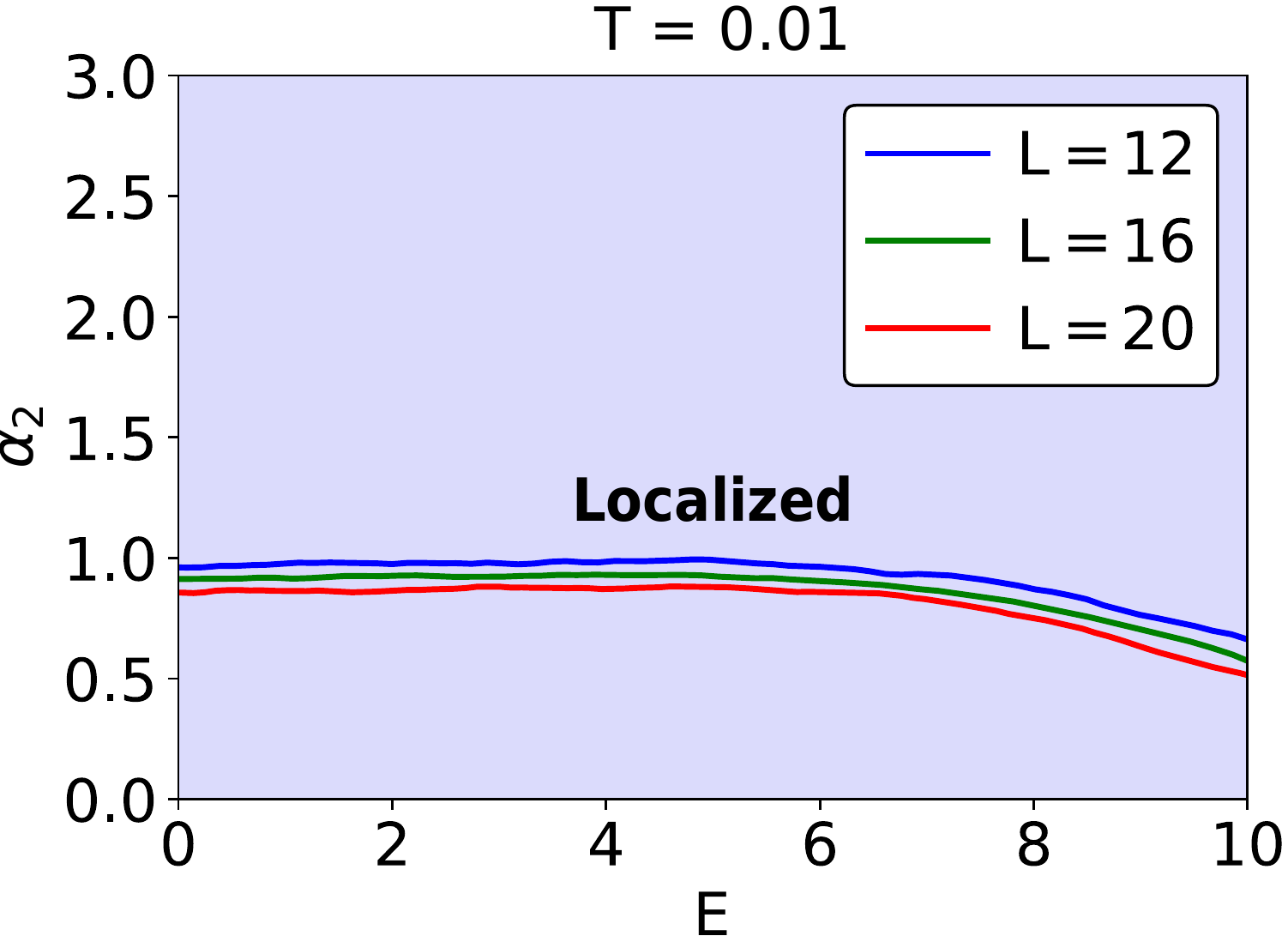}
\includegraphics[scale=0.25]{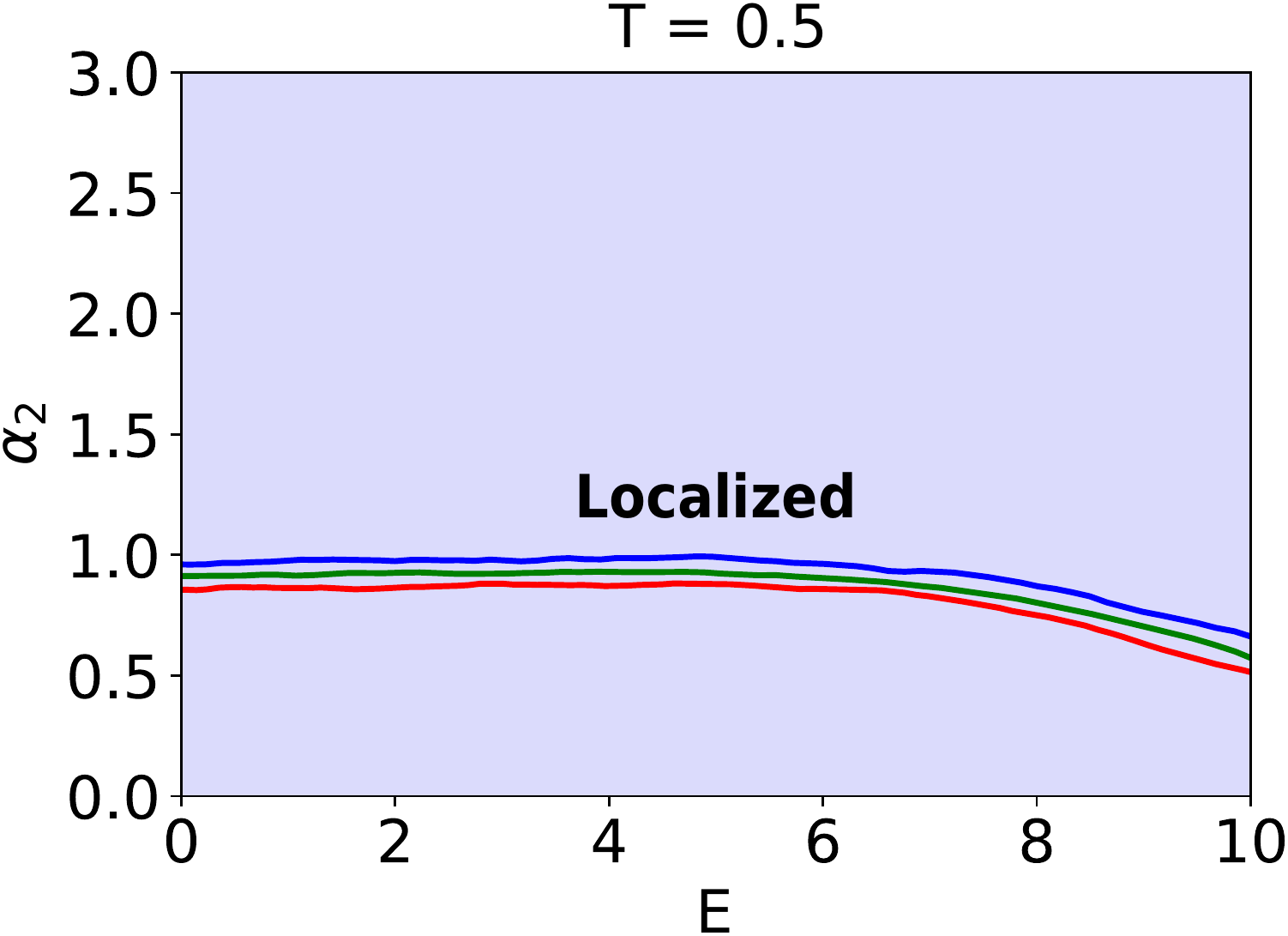}
  \caption{Scaling behavior of $\alpha_2$ at finite temperatures of T=0.01 and T=0.5 for the type I insulator ($U=0.1<U_{c1}$). There do not appear any insulator to metal phase transitions, increasing temperatures up to the bandwidth. In this respect we identify the type I insulator with an Anderson insulating phase.}
 \label{fig:MBL_AI}
\end{figure}

\begin{figure}
\includegraphics[scale=0.25]{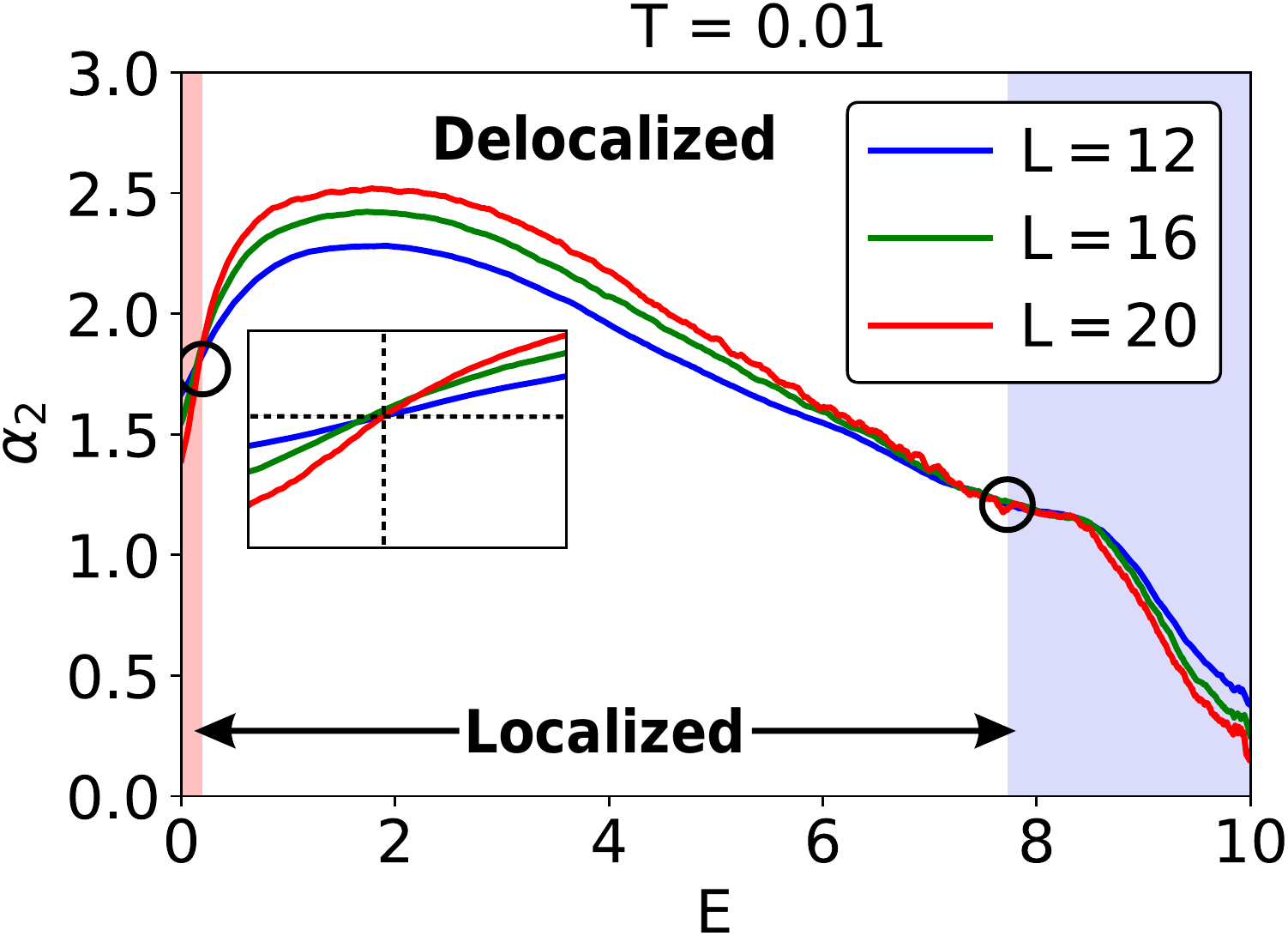}
\includegraphics[scale=0.25]{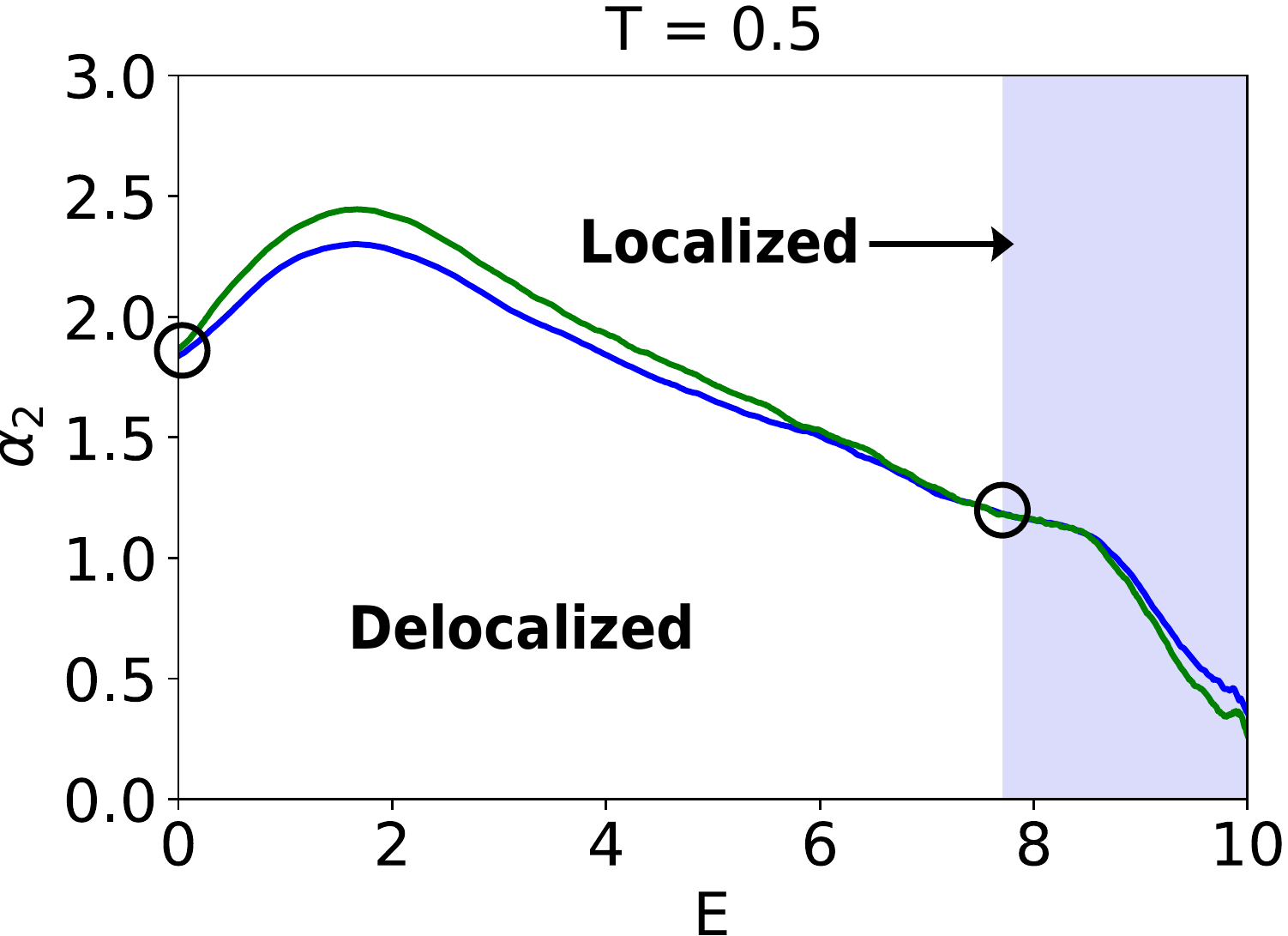}
  \caption{Scaling behavior of $\alpha_2$ at finite temperatures of $T=0.01$ and $T=0.5$ for the type II insulator ($U=5>U_{c2}$). The type I mobility edge at a high energy does not change, increasing temperatures. On the other hand, the type II mobility edge near the Fermi energy moves toward the Fermi energy and disappear eventually, increasing temperature further. As a result, there does appear an insulator to metal transition at a critical temperature near the Fermi energy. In this respect we identify the type II insulator with an MBL insulating phase.}
 \label{fig:MBL_U5_MFA}
\end{figure}

\begin{figure}
\includegraphics[scale=0.25]{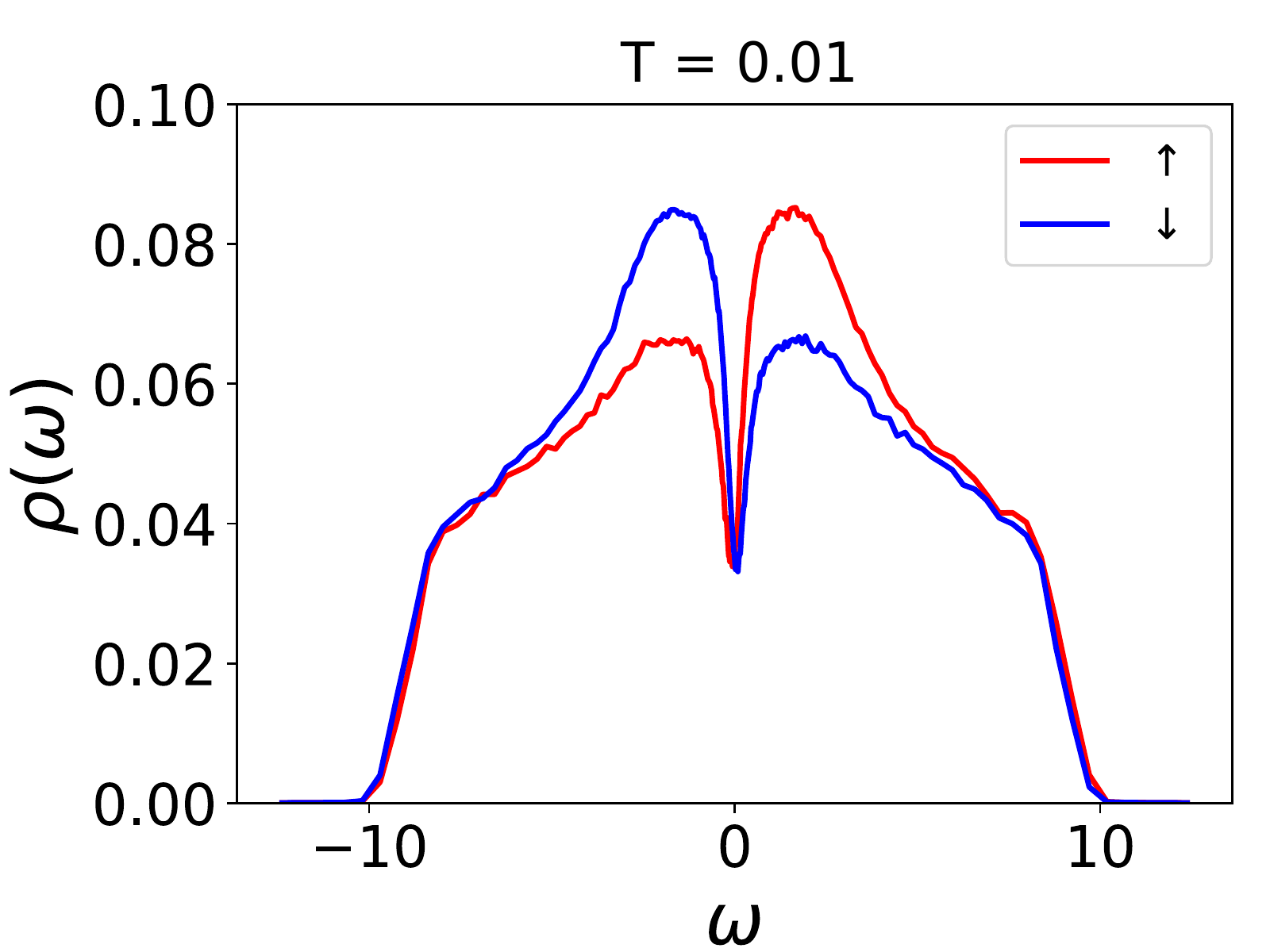}
\includegraphics[scale=0.25]{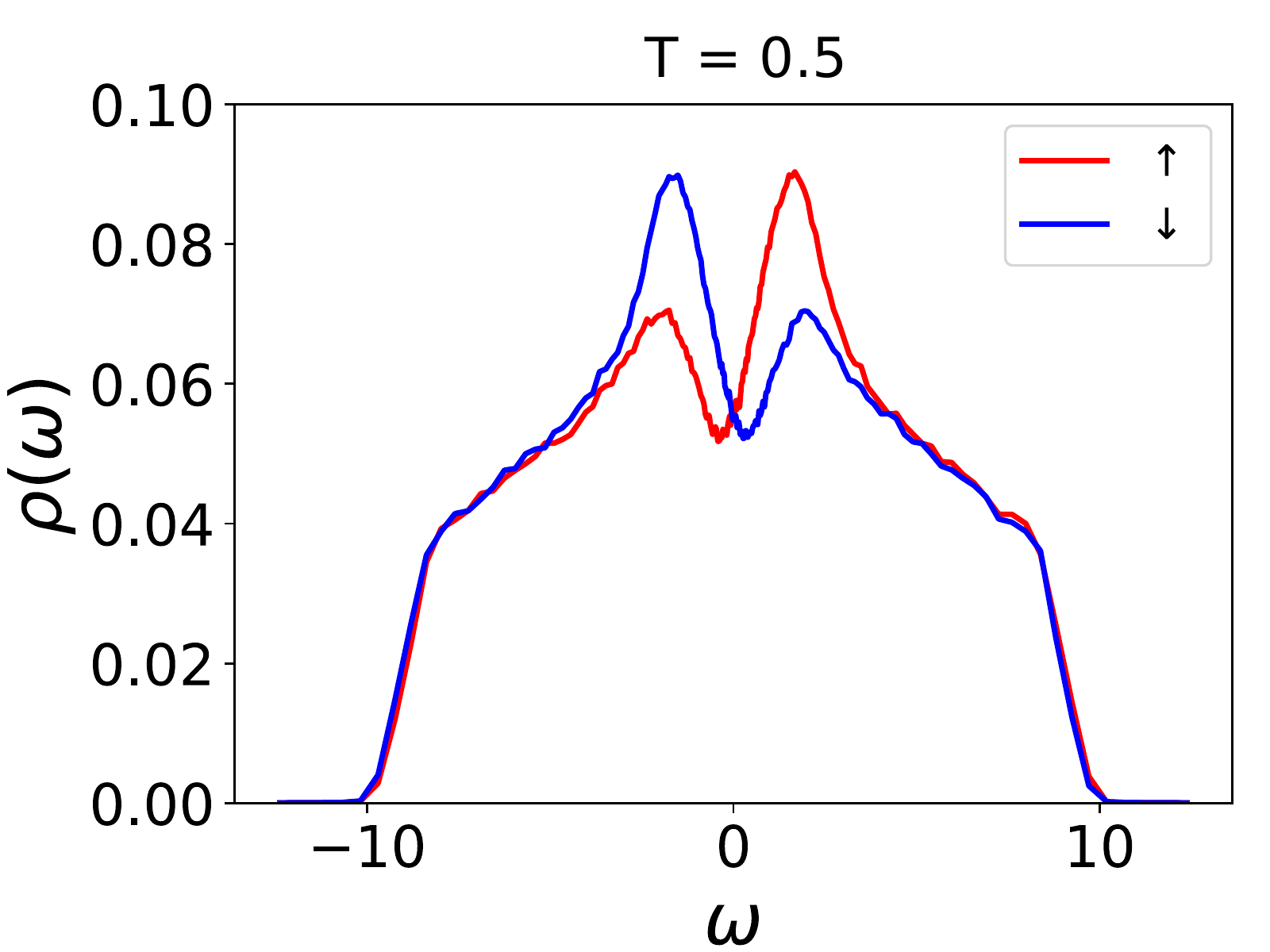}
  \caption{Density of states at finite temperatures of $T=0.01$ and $T=0.5$ for the type II insulator ($U=5>U_{c2}$). The red (blue) line represents the density of states for spin $\uparrow$ ($\downarrow$) electrons. The pseudogap is filled at finite temperatures. Enhancement of the density of states near the Fermi energy is responsible for the insulator to metal transition at a critical temperature from the type II insulator.}
 \label{fig:MBL_U5_DOS}
\end{figure}

To distinguish the type II insulator from the type I insulator, we investigate how the multifractal exponent $\alpha_{2}$ evolves as a function of temperature.
%
%
Figure \ref{fig:MBL_AI} shows the development of the scaling behavior of the IPR ($\alpha_2$) as a function of temperature for the type I insulator ($U=0.1<U_{c1}$). The localized state remains to be stable even with interactions. No quantitative changes are observed for the case of the type I insulator. On the other hand, the type II insulator ($U=5>U_{c2}$) shows enhancement of the density of states around the Fermi energy due to excitations at finite temperatures. See Figs. \ref{fig:MBL_U5_DOS} (a) and (b). As a result, the pseudogap is filled around the critical temperature ($T_c \approx 0.5$). Figures \ref{fig:MBL_U5_MFA} (a) and (b) show scaling of the IPR ($\alpha_2$) in the type II insulator as a function of temperature. Near the zero temperature ($T=0.01$), there exist low- and high-energy mobility edges. We point out that the high-energy mobility edge is little affected by finite-temperature excitations.
%
%
On the other hand, we observe that the low-energy mobility edge starts to shift to the Fermi level due to enhancement of the density of states around the Fermi energy. This low-energy mobility edge coincides with the Fermi level from the critical temperature ($T_{c} \approx 0.5$). When $T>T_c$, the low-energy mobility edge disappears and the diffusive nature comes to appear. See Fig \ref{fig:MBL_U5_MFA} (b).

\subsection{Area law of entanglement entropy and beyond}

To understand the properties of type-I and type-II insulators further, we obtain the entanglement entropy for both systems. We consider subregions of the linear size $l=4, 5 , 6 $, and 7, then analyze scaling of the entanglement entropy. Here, the quantity $x$ is defined as
\begin{align}
    x_{ll'}=\frac{\log (S_E(l)/S_E(l'))}{\log ({l/l'})}
\end{align}
We obtain $x \approx 2.15$ for both insulators, which indicates the area law of the entanglement entropy. However, larger values of the entanglement entropy imply that the type-II insulator is more entangled than the type-I insulator. Since type-II has delocalized states at high energies, this result is consistent with our intuition.

\begin{table}[]
\centering
\begin{tabular}{|c|c|c|c|c|c|}
\cline{1-6}
           & $l=4$  & $l=5$  &$l=6$  &$l=7$ & $\la x \ra$ \\\cline{1-6}
     Type-I $S_E(l)$ & $5.23$ & $8.61$& $12.56$ &$17.52$&$2.15$      \\\cline{1-6}
     Type-II $S_E(l)$ & $7.56$ & $12.37$ & $18.15$ &$25.2$&$2.15$   \\\cline{1-6}
\end{tabular}
\caption{Entanglement entropy for Type-I and Type-II insulators as a function of the subsystem size $l^3$. See the text for the definition of $x$. Although $ \la x \ra = 2.15$ confirms that both insulators follow the area law of the entanglement entropy, the type II insulating phase shows stronger entanglement than the type I.}
\label{table1}
\end{table}

\section{Conclusion}

In this study, we uncovered two types of insulting phases and their quantum phase transitions via an intermediate metallic state as a function of interactions and single-particle eigen-energies. See the phase diagram of Fig. \ref{fig:phase1}. In particular, we identified two types of mobility edges, referred to as type I and type II. Spin degrees of freedom turn out to play a central role in these dynamical phase transitions: Enhancement of the density of states via spin polarization near the Fermi energy gives rise to the type I mobility edge while the formation of a pseudogap is responsible for the type II mobility edge. The multifractal spectrum of the type I mobility edge turns out to be smoothly connected with that of the Anderson transition in the absence of electron correlations. This type I insulating phase survives up to the temperature of the order of the bandwidth. As a result, the type I insulating state was identified with a paramagnetic Anderson insulating phase. On the other hand, the multifractal spectrum of the type II mobility edge describes another quantum phase transition between a ferromagnetic insulator and a ferromagnetic metal, which belongs to a different universality class from the Anderson transition without correlations. First of all, we revealed that there appears a phase transition at a critical temperature from a ferromagnetic insulator to a ferromagnetic metal, increasing temperatures. In this respect we concluded that the type II insulating phase may be identified with an MBL insulating state.

%
%

Before closing our discussions, we give several remarks. Consider the case without spin polarization. As mentioned previously \cite{Comments_no_polarization}, enhancement of the density of states is also allowed in the Hubbard-type model without spin polarization. In this respect it is natural to expect that the first quantum phase transition from a paramagnetic Anderson insulating phase to a paramagnetic metallic state would occur. However, it is not clear what happens in appearance of the pseudogap without spin polarization, increasing interactions further. A Mott gap may arise in this spinless case, where the Hartree-Fock-Anderson simulation method does not apply. One can also ask what happens if we take into account effective interactions beyond the Hubbard type. Although we assumed effective local interactions in the insulating regime of parameters, it would be more natural to consider Coulomb interactions. Previously, we investigated the role of Coulomb interactions in Anderson localization of spinless fermions, where the weak-localization regime has been considered \cite{Two_Mobility_Edges,Half_metals}. Extending this previous study to the strong disorder regime, it would be an interesting future direction of research to investigate whether the intermediate metallic state remains stabilized or not.

\begin{acknowledgments}
K.-S. Kim was supported by the Ministry of Education, Science, and Technology (NRF-2021R1A2C1006453 and NRF-2021R1A4A3029839) of the National Research Foundation of Korea (NRF) and by TJ Park Science Fellowship of the POSCO TJ Park Foundation.
\end{acknowledgments}


\begin{thebibliography}{99}
\bibitem{Anderson_AL_Original} P. W. Anderson, \textit{Absence of Diffusion in Certain Random Lattices}, Phys. Rev. {\bf 109}, 1492 (1958).
\bibitem{AL_Disorder_Interaction} B. L. Altshuer, A. G. Aronov, A. L. Efros, and M. Pollak, \textit{Electron-electron Interactions in Disordered Systems}, (Elsevier, Amsterdam, 1985).
\bibitem{AL_Thermalization_I} L. Fleishman and P. W. Anderson, \textit{Interactions and the Anderson transition}, Phys. Rev. B {\bf 21}, 2366 (1980).
\bibitem{AL_Thermalization_II} T. V. Shahbazyan and M. E. Raikh, \textit{Surface plasmon in a two-dimensional Anderson insulator with interactions}, Phys. Rev. B {\bf 53}, 7299 (1996).
\bibitem{AL_Thermalization_III} V. I. Kozub, S. D. Baranovskii, and I. Shlimak, \textit{Fluctuation-stimulated variable-range hopping}, Solid State Commun. {\bf 113}, 587 (2000).
\bibitem{AL_Thermalization_IV} B. L. Altshuler, Y. Gefen, A. Kamenev, and L. S. Levitov, \textit{Quasiparticle Lifetime in a Finite System: A Nonperturbative Approach} Phys. Rev. Lett. {\bf 78}, 2803 (1997).
\bibitem{MBL_Confirmed_I} D. Basko, I. L. Aleiner, and B. L. Altshuler, \textit{Metal.insulator transition in a weakly interacting many-electron system with localized single-particle states}, Ann. Phys. {\bf 321}, 1126 (2006).
\bibitem{MBL_Confirmed_II} I. Gornyi, A. Mirlin, and D. Polyakov, \textit{Interacting Electrons in Disordered Wires: Anderson Localization and Low-T  Transport}, Phys. Rev. Lett. {\bf 95}, 206603 (2005).
\bibitem{MBL_Review_I} E. Altman and R. Vosk, \textit{Universal dynamics and renormalization in many-body-localized systems}, Annu. Rev. Condens. Matter Phys. {\bf 6}, 383 (2015).
\bibitem{MBL_Review_II} R. Nandkishore and D. A. Huse, \textit{Many-body localization and thermalization in quantum statistical mechanics}, Annu. Rev. Condens. Matter Phys. {\bf 6}, 15 (2015).
\bibitem{MBL_Review_III} D. A. Abanin, E. Altman, I. Bloch, and M. Serbyn, \textit{Colloquium: Many-body localization, thermalization, and entanglement}, Rev. Mod. Phys. {\bf 91}, 021001 (2019).
\bibitem{MBL_Proof_1D} J. Z. Imbrie, \textit{On many-body localization for quantum spin chains}, J. Stat. Phys. {\bf 163}, 998 (2016).
\bibitem{Altaman_Review} E. Altman, \textit{Many-body localization and quantum thermalization}, Nat. Phys. {\bf 14}, 979 (2018).
\bibitem{MBL_Selected_Topics} F. Alet and N. Laflorencie, \textit{Many-body localization: An introduction and selected topics}, C. R. Physique, {\bf 19}, 498 (2018).
\bibitem{MBL_QPT_RG_I} R. Vosk, D. A. Huse, and E. Altman, \textit{Theory of the many-body localization transition in one-dimensional systems}, Phys. Rev. X {\bf 5}, 031032 (2015).
\bibitem{MBL_QPT_RG_II} A. C. Potter, R. Vasseur, and S. A. Parameswaran, \textit{Universal properties of many-body delocalization transitions}, Phys. Rev. X {\bf 5}, 031033 (2015).
\bibitem{MBL_QPT_RG_III} Y. Bar Lev, G. Cohen, and D. R. Reichman, \textit{Absence of diffusion in an interacting system of spinless fermions on a one-dimensional disordered lattice}, Phys. Rev. Lett. {\bf 114}, 100601 (2015).
\bibitem{MBL_QPT_RG_IV} K. Agarwal, S. Gopalakrishnan, M. Knap, M. Muller, and E. Demler, \textit{Anomalous diffusion and Griffiths effects near the many-body localization transition}, Phys. Rev. Lett. {\bf 114}, 160401 (2015).
\bibitem{MBL_QPT_RG_V} M. Znidaric, A. Scardicchio, and V. K. Varma, \textit{Diffusive and subdiffusive spin transport in the ergodic phase of a many-body localizable system}, Phys. Rev. Lett. {\bf 117}, 040601 (2016).
\bibitem{Finkelstein_RG} A. Punnoose and A. M. Finkelstein, \textit{Metal-Insulator Transition in Disordered Two-Dimensional Electron Systems}, Science {\bf 310}, 289 (2005); S. Anissimova, S. V. Kravchenko, A. Punnoose, A. M. Finkelstein, and T. M. Klapwijk, \textit{Metal-Insulator Transition in Disordered Two-Dimensional Electron Systems}, Nat. Phys. {\bf 3}, 707 (2007).
\bibitem{Two_Mobility_Edges} Hyun-Jung Lee and Ki-Seok Kim, \textit{Hartree-Fock study of the Anderson metal-insulator transition in the presence of Coulomb interaction: Two types of mobility edges and their multifractal scaling exponents}, Phys. Rev. B {\bf 97}, 155105 (2018).
\bibitem{Half_metals} Kyung-Yong Park, Hyun-Jung Lee, and Ki-Seok Kim, \textit{Half metals at intermediate energy scales in Anderson-type insulators}, Phys. Rev. B {\bf104}, 054206 (2021).
\bibitem{IPR_Review} F. Evers and A. D. Mirlin, \textit{Anderson transitions}, Rev. Mod. Phys. {\bf 80}, 1355 (2008).
\bibitem{CPA_plus_MFT_AL} P. Henseler, J. Kroha, and B. Shapiro, \textit{Self-consistent study of Anderson localization in the Anderson-Hubbard model in two and three dimensions}, Phys. Rev. B {\bf 78}, 235116 (2008).
\bibitem{Level_Statistics_Review} A. D. Mirlin, \textit{Statistics of energy levels and eigenfunctions in disordered systems}, Phys. Rep. {\bf 326}, 259 (2000).
\bibitem{Stoner_Original} E. C. Stoner, \textit{Ferromagnetism}, Rep. Prog. Phys. 11, 43 (1947).
\bibitem{Stoner_Instability_Textbook} N. Nagaosa, \textit{Quantum Field Theory in Strongly Correlated Electronic Systems} (Springer-Verlag, New York, 1999).
\bibitem{AMIT_Critical_Disorder_Strength} K. Slevin and T. Ohtsuki, \textit{Corrections to Scaling at the Anderson Transition}, Phys. Rev. Lett. {\bf 82}, 382 (1999).
\bibitem{Chhabra1989} A. Chhabra and R. V. Jensen, \textit{Direct determination of the $f(\alpha)$ singularity spectrum}, Phys. Rev. Lett. {\bf 62}, 1327 (1989).
\bibitem{Janssen1994} M. Janssen, \textit{Mutifractal Analysis of Broadly Distributed Observables at Criticality}, Int. J. Mod. Phys. B {\bf 8}, 943 (1994).
\bibitem{ChungPeschel} M.-C. Chung and I. Peschel, \textit{Density-Matrix Spectra of Solvable Fermionic Systems}, Phys. Rev. B {\bf64}, 064412 (2001).
\bibitem{Peschel} I. Peschel, \textit{Calculation of reduced density matrices from correlation functions}, J. Phys. A: Math. Gen. 36 L205 (2003).
\bibitem{ChungHenley} S.-A. Cheong and C. L. Henley, \textit{Many-body density matrices for free fermions}, Phys. Rev. B {\bf 69}, 075111 (2004).

\bibitem{Comments_no_polarization} We point out that enhancement of the density of states is also observed in the spinless case \cite{CPA_plus_MFT_AL}. We give our critical discussons on this aspect in conclusion.

%

%
%
%
%
%
%

%
%

%
\end{thebibliography}
\end{document}